\documentstyle[12pt]{article}
\begin{document}

\noindent Stockholm\\
July 2007\\

\vspace{12mm}

\begin{center}

{\Large A CURIOUS GEOMETRICAL FACT ABOUT} 

\vspace{10mm}

{\Large ENTANGLEMENT}\footnote{Talk at the 
V\"axj\"o conference on Quantum Theory: Reconsideration of 
Foundations, 

June 2007.} 

\vspace{15mm}

{\large Ingemar Bengtsson}\footnote{Email address: ingemar@physto.se. 
Supported by VR.}

\vspace{10mm}

{\sl Stockholm University, Alba Nova\\
Fysikum\\
S-106 91 Stockholm, Sweden}

\vspace{15mm}

{\bf Abstract}

\end{center}

\vspace{5mm}

\noindent I sketch how the set of pure quantum states forms a phase space, and then 
point out a curiousity concerning maximally entangled pure states: they form a 
minimal Lagrangian submanifold of the set of all pure states. I suggest 
that this curiousity should have an interesting physical interpretation.

\newpage

{\bf 1. From classical to quantum states}

\vspace{5mm}

\noindent I will state a curious fact concerning maximally entangled pure states 
of a bipartite quantum system \cite{BBZ}. I do not know any physical interpretation 
of it, but I do expect that there is one. Before I can state 
my fact, I will have to give a bird's eye view of some aspects of 
quantum mechanics---aspects that may be worth recalling anyway, in a 
reconsideration of the foundations of quantum theory. 

We begin at the beginning. A classical state---in classical 
probability theory---is a probability distribution, that is a set of 
$N$ numbers $p_i$ such that 

\begin{equation} p_i \geq 0 \ , \hspace{10mm} \sum_{i=1}^N p_i = 1 \ . 
\end{equation}

\noindent The set of classical states forms a simplex; a triangle if 
$N = 3$, which is a good case to choose as an illustrative example. 
If you want to introduce dynamics on a probability simplex, the options are 
rather limited. You can use stochastic maps, tending to shrink the simplex 
towards some fixed point, and then 
combine such maps into Markov chains. If we also require the stochastic 
maps to take pure states into pure states, we only have permutations of 
the corners to play with. 

The set of pure states is isomorphic to the sample space. To define a 
symplectic structure on sample space, it needs to be a continuous space---the 
discrete space forming the corners of a finite dimensional 
simplex simply will not do. To get classical mechanics, with Hamiltonians 
and symplectic forms, it is necessary to go to the infinite dimensional 
case. (I am aware that this statement must be qualified \cite{Subhash}.)

To generalize the classical case, we first rewrite the definitions in terms 
of diagonal matrices. Thus a state $P$ is a matrix 

\begin{equation} P = \left( \begin{array}{lll} p_1 & 0 & 0 \\ 0 & p_2 & 0 \\ 
0 & 0 & p_3 \end{array} \right) \ , \hspace{5mm} P \geq 0 \ , \hspace{3mm} 
\mbox{Tr}P = 1 \ . \end{equation}

\noindent A random variable $A$ is an otherwise unrestricted diagonal matrix, 

\begin{equation} A = \left( \begin{array}{lll} a_1 & 0 & 0 \\ 0 & a_2 & 0 \\ 
0 & 0 & a_3 \end{array} \right) \ . \end{equation}

\noindent The expectation value of a random variable, given a state, is then 

\begin{equation} \langle A\rangle = \mbox{Tr}PA \ . \label{5} \end{equation}

There is a fairly obvious generalization of all this. We replace the diagonal matrices 
with ``diagonalizable'', that is Hermitian, matrices. This still leaves it open 
whether they should be real or complex matrices, but I will soon argue that 
complex numbers are the preferred choice, so let us make it right away. A 
state is now a positive (and therefore Hermitian) matrix, 

\begin{equation} \rho = \rho^\dagger \geq 0 \ , \hspace{9mm} \mbox{Tr}\rho = 
1 \ . \end{equation}

\noindent Such a state is known as a density matrix. A random variable is an 
otherwise unrestricted Hermitian matrix, 

\begin{equation} A = A^{\dagger} \ . \end{equation}

\noindent For expectation values, we keep the classical formula (\ref{5}). 

The quantum generalization is a significant one. The random variables, and the 
states, now belong to a non-commutative algebra. The $N-1$ dimensional classical state 
space is turned into an $N^2 - 1 = (N+1)(N-1)$ dimensional one. This set forms an 
interesting convex body, which can be understood by observing that a 
general density matrix can be obtained from a diagonal one by means of 
a unitary transformation 

\begin{equation} \rho \rightarrow U\rho U^\dagger \ . \end{equation}

\noindent It is only the $SU(N)$ subgroup that acts effectively on the 
states. This is a subgroup of $SO(N^2-1)$, so the convex body of density 
matrix is obtained by performing quite special rotations of the classical 
simplex. This gives the convex body a subtle structure, but our concern 
here is with its pure states. 

Pure quantum states are density matrices of rank one,

\begin{equation} \rho = \frac{\psi^\alpha \psi_\beta}{\psi \cdot 
\bar{\psi}} \ , \hspace{5mm} \psi \cdot \bar{\psi} \equiv \psi^\alpha
\bar{\psi}_\alpha = \langle 
\psi |\psi \rangle \ , \hspace{5mm} 1 \leq \alpha, \beta \leq N \ . 
\end{equation}

\noindent Equivalently, they are vectors in an $N$ dimensional Hilbert 
space up to normalization and phase, that is to say they can be regarded 
as points in complex projective space ${\bf CP}^{N-1}$. The real dimension 
of this space is $2(N-1)$. Hence the pure states in quantum mechanics form 
a continuous manifold already in the finite dimensional case. 

\vspace{1cm}
 
{\bf 2. Quantum mechanics}

\vspace{5mm}

\noindent Because we decided to work with complex numbers, we can now justify the 
name ``quantum mechanics''. This happens because, as noted by Strocchi 
\cite{Strocchi} (and later by others \cite{Kibble,Gibbons,Hughston,Ashtekar}), 
the Hilbert space scalar product conceals a symplectic form. 

To see this, split the complex vectors into real and imaginary parts, 
and reassemble them into real $2N$ dimensional vectors:

\begin{equation} \psi^\alpha = x^\alpha + iy^\alpha \ , \hspace{10mm} 
X^I = \left( \begin{array}{c} x^\alpha \\ y^\alpha \end{array} 
\right) \ , \hspace{4mm} 1 \leq I, J \leq 2N \ . \end{equation}

\noindent Similarly, Hermitian operators may be split into real 
symmetric and imaginary anti-symmetric parts. The scalar product 
becomes 

\begin{equation} X\cdot Y = X^Ig_{IJ}Y^J + iX^I\Omega_{IJ}Y^J \ , 
\end{equation}

\noindent where 

\begin{equation} g_{IJ} = \left( \begin{array}{cc} {\bf 1} & 0 \\ 
0 & {\bf 1} \end{array} \right) \hspace{8mm} 
\Omega_{IJ} = \left( \begin{array}{cc} 0 & {\bf 1} \\ 
- {\bf 1} & 0 \end{array} \right) \ . \end{equation}

\noindent Therefore $g_{IJ}$ will serve as a metric on state space, while 
$\Omega_{IJ}$ is an anti-symmetric symplectic form. 

Recall that a symplectic form on a vector space is simply a non-degenerate 
anti-symmetric matrix. If $\Omega^{IJ}$ denotes its inverse, one can use it to 
define a Poisson bracket between any two functions on phase space, as 

\begin{equation} \{ f(X), g(X)\} = \Omega^{IJ}\partial_If(X)\partial_Jg(X) 
\ . \end{equation}

\noindent If the symplectic form itself depends on $X^I$, an extra 
condition---that the form be closed---must be imposed in order to ensure that 
the Jacobi identities hold.

The point we are driving at is that, once these definitions are in place, 
the Schr\"odinger equation takes precisely the form that time evolution 
always takes in classical Hamiltonian mechanics: 

\begin{equation}i\partial_t\psi^\alpha = H^\alpha_{\ \beta}\psi^\beta 
\hspace{4mm} \Leftrightarrow \hspace{4mm} \dot{X}^I = \Omega^{IJ}\partial_J
\langle H \rangle = \{ X^I, \langle H\rangle \} \ . \end{equation}

\noindent It is just that the Hamiltonian function takes a quite special form: 

\begin{equation} \langle H \rangle = \langle \psi |H| \psi \rangle \ . 
\end{equation}

\noindent Hence the name quantum mechanics is indeed justified: the space 
of its pure states is rich enough to support time evolution of Hamiltonian 
form. We have not identified classical and quantum 
mechanics, since we admit Hamiltonians of a very special 
form only. On inspection, one finds that quantum mechanics admits precisely 
those Hamiltonian flows that preserve the metric which is also present on phase 
space.  

In classical theory, one needs a continuous sample space to support Hamiltonian 
mechanics. In quantum theory Hamiltonian mechanics is always there, because 
one uses a continuous set of discrete (or continuous) sample spaces.

We note in passing that nothing like this would work over the real 
numbers. Real projective space is not a symplectic manifold.

For the record, we will need the symplectic form also on complex projective 
space. On ${\bf CP}^{N-1}$ we have the Fubini-Study metric 

\begin{equation} ds^2 = \frac{\psi \cdot \bar{\psi} 
d\psi \cdot d\bar{\psi} - d\psi \cdot \bar{\psi} 
\psi \cdot d\bar{\psi}}{\psi \cdot \bar{\psi} \psi \cdot \bar{\psi}} \ , 
\end{equation}

\noindent and its relative, the symplectic form 

\begin{equation} \Omega = i\frac{\psi \cdot \bar{\psi} 
d\psi \cdot \wedge d\bar{\psi} - d\psi \cdot \bar{\psi} 
\wedge \psi \cdot d\bar{\psi}}{\psi \cdot \bar{\psi} \psi \cdot \bar{\psi}}\ . 
\label{symp} \end{equation}

\noindent Basically this is what we had before, rewritten to be invariant 
under the transformation $\psi^\alpha \rightarrow z \psi^\alpha$, where 
$z$ is an arbitrary non-zero complex number.

\vspace{1cm}

{\bf 3. Composite systems}

\vspace{5mm}

\noindent We are now half way to our fact. Before we get there, we must consider composite 
systems, described by means of the Hilbert space ${\cal H}^{N^2} = 
{\cal H}^N \otimes {\cal H}^N$. For this purpose, it is helpful to introduce 
a notation that regards vectors as square arrays of numbers. Thus we write 

\begin{equation} |\Psi \rangle = \psi^\alpha |e_\alpha\rangle = 
\frac{1}{\sqrt{N}}\Gamma^{ij}|e_i\rangle |e_j\rangle \ , \hspace{5mm} 
1 \leq \alpha \leq N^2 \ , \hspace{3mm} 1 \leq i, j \leq N \ . \end{equation}

\noindent Although $\Gamma^{ij}$ is primarily a square array, not a matrix, 
the terminology of matrix theory applies to it. If the matrix is of rank one, 
the state is a product state: 

\begin{equation} |\Psi\rangle = \phi^i\lambda^j|e_i\rangle |e_j\rangle = 
|\phi\rangle |\lambda \rangle \ . \end{equation}

\noindent Such states are also known as separable. Geometrically, the set 
of separable states are embedded in ${\bf CP}^{N^2-1}$ through the Segre 
embedding 

\begin{equation} {\bf CP}^N\times {\bf CP}^N \rightarrow {\bf CP}^{N^2-1} 
\ . \end{equation}

\noindent They form an orbit under the group of local unitaries, $SU(N)
\times SU(N)$, which is a subgroup of $SU(N^2)$. 

Once vectors get two indices, density matrices get four. The rank one 
density matrix describing a pure state is  

\begin{equation} \rho^{ij}_{\ \ kl} = \psi^\alpha \bar{\psi}_\beta = \frac{1}{N} 
\Gamma^{ij}\Gamma_{kl}^* \ . \end{equation}  

\noindent A basic point is this: if we control operators of the form $A\otimes {\bf 1}$ 
only, then the relevant state is given by the partial trace 

\begin{equation} \mbox{Tr}_2 \rho = \frac{1}{N}\sum_{j= 1}^N\Gamma^{ij}\Gamma_{kj}^* = 
\frac{1}{N}(\Gamma \Gamma^\dagger )^i_{\ k} \ . \end{equation}

\noindent If the original state is separable, this is still a pure state on 
${\cal H}^N$. On the other hand, if  

\begin{equation} \Gamma \Gamma^\dagger = {\bf 1} \ , \end{equation}

\noindent then $\Gamma \in U(N)$, 
the reduced state is maximally mixed, and the original state is 
said to be maximally entangled---all the information in the original state is 
in (non-classical) correlations. 

Projectively the phase is irrelevant, and the set of maximally entangled states 
in ${\cal H}^N\otimes {\cal H}^N$ is isomorphic to 

\begin{equation} SU(N)/Z_N \in {\bf CP}^{N^2 -1} \ . \end{equation}

\noindent This is a group manifold, and as such it is again an orbit under 
$SU(N)\times SU(N)$. But precisely what kind of submanifold is it?

\vspace{1cm}

{\bf 4. The curious fact}

\vspace{5mm}

\noindent Observe that 

\begin{equation} \mbox{dim}\left[ SU(N)/Z_N\right] = N^2 - 1 = \frac{1}{2}\mbox{dim}
\left[ {\bf CP}^{N^2-1}\right] \ . \end{equation}

\noindent This should make you suspicious. If we insert a state vector given 
by a unitary matrix in the $N^2$ dimensional version of eq. (\ref{symp}), 
we find \cite{BBZ} that 

\begin{equation} \Omega_{|_{SU(N)/Z_N}} = 0 \ . \end{equation}

\noindent The dimension of the set of maximally entangled state is one half 
of the dimension of the set of all pure states, and the symplectic form 
vanishes when restricted to this submanifold. A submanifold having these 
two properties is called a Lagrangian submanifold. To a physicist, Lagrangian 
submanifolds are known as configuration spaces. In a phase space spanned by 
$q$s and $p$s, the subspace spanned by the $q$s is Lagrangian. 

Considered as a fact about differential geometry, our example is well 
known to mathematicians \cite{Wang}. They also know that this particular 
example is a minimal submanifold---if you wiggle it, its volume grows. In a technical 
sense, its two properties make our submanifold into a special submanifold of 
a K\"ahler space. 

We have arrived at our curious fact: {\it The set of maximally entangled pure 
states is a Lagrangian and minimal submanifold of the set of all pure states.}
From a physical point of view, why? I do not know, but I imagine that 
there should be a reason, perhaps one that can be phrased in control theoretic 
terms. I did offer a free dinner to 
any participant who would tell me, but there were no takers. The offer still stands. 

\newpage

{\bf Acknowledgements}

\vspace{5mm}

\noindent I thank Johan Br\"annlund and Karol \.Zyczkowski for collaboration, 
Robert Bryant and Vladimir Manko for supplying references, Ansar 
Fayyazuddin for discussions, and Andrei 
Khrennikov for being my host in Sm\aa land.

\end{document}